\documentclass[fleqn,usenatbib]{mnras}

\usepackage{newtxtext,newtxmath}

\usepackage[T1]{fontenc}

\DeclareRobustCommand{\VAN}[3]{#2}
\let\VANthebibliography\thebibliography
\def\thebibliography{\DeclareRobustCommand{\VAN}[3]{##3}\VANthebibliography}

\usepackage{graphicx}	
\usepackage{amsmath}	
\usepackage{subfigure}

\newcommand{\mystar}{HATS-46}

\newcommand{\python}{\ensuremath{\textsc{Python}}}

\title[Transmission spectrum of HATS-46\,b]{
LRG-BEASTS: Evidence for clouds in the transmission spectrum of HATS-46 b}

\author[E. Ahrer et al.]{
E. Ahrer$^{1,2}$\thanks{E-mail: eva-maria.ahrer@warwick.ac.uk},
P. J. Wheatley$^{1,2}$\thanks{E-mail: p.j.wheatley@warwick.ac.uk},
 S. Gandhi$^{3,1,2}$, J. Kirk$^{4,5}$, G. W. King$^{6,1,2}$,
T. Louden$^{1,2}$ and L. Welbanks$^{7}$\thanks{NHFP Sagan Fellow}
\\
$^{1}$Centre for Exoplanets and Habitability, University of Warwick, Gibbet Hill Road, CV4 7AL Coventry, UK\\
$^{2}$Department of Physics, University of Warwick, Gibbet Hill Road, CV4 7AL Coventry, UK\\
$^{3}$Leiden Observatory, Leiden University, Postbus 9513, 2300 RA Leiden, The Netherlands\\
$^{4}$Center for Astrophysics | Harvard \& Smithsonian, 60 Garden Street, Cambridge, MA 02138, USA\\
$^{5}$Department of Physics, Imperial College London, Prince Consort Road, London, SW7 2AZ, UK\\
$^{6}$Department of Astronomy, University of Michigan, Ann Arbor, MI 48109, USA\\
$^{7}$School of Earth \& Space Exploration, Arizona State University, Tempe, AZ, 85257, USA\\
}

\date{Accepted XXX. Received YYY; in original form ZZZ}

\pubyear{2023}

\begin{document}
\label{firstpage}
\pagerange{\pageref{firstpage}--\pageref{lastpage}}
\maketitle

\begin{abstract}
  We have performed low-resolution ground-based 
  spectroscopy of HATS-46\,b in transmission, using the EFOSC2 instrument on the ESO New Technology Telescope (NTT). HATS-46\,b is a highly-inflated exoplanet that is a prime target for transmission spectroscopy, having a Jupiter-like radius (0.95 R$_\textrm{Jup}$) but a much lower mass (0.16 M$_\textrm{Jup}$). 
  It orbits a G-type star with a 4.7\,d period, giving an equilibrium temperature of 1100\,K.  
  We observed one transit of HATS-46\,b with the NTT, with the time-series spectra covering a wavelength range of 3900 - 9000 \AA\ at a resolution of $R \sim 380$. We achieved a remarkably precise transmission spectrum of 1.03 $\times$ photon noise, with a median uncertainty of $357$\,ppm for $\sim 200$\,\AA\ wide bins, despite the relative faintness of the host star with $V_{\mathrm{mag}} = 13.6$. The transmission spectrum does not show strong absorption features and retrievals favour a cloudy model, ruling out a clear atmosphere with $3.0\sigma$ confidence. We  also place a conservative upper limit on the sodium abundance under the alternative scenario of a clear atmosphere. 
  This is the eighth planet in the LRG-BEASTS survey, which uses 4\,m-class telescopes such as the NTT to obtain low-resolution transmission spectra of hot Jupiters with precisions of around one atmospheric scale height.
\end{abstract}




\section{Introduction}

The study of transit depth versus wavelength, or {\em transmission spectroscopy}, is an essential 
method to characterise the atmospheres of transiting exoplanets with both ground- and space-based telescopes \citep[e.g.][]{Charbonneau2002DetectionAtmosphere, Snellen2008Ground-based209458b,Bean2010A1214b, Stevenson2014Transmissionm,  Sing2016ADepletion,May2018MOPSSWASP-52b, Weaver2021ACCESS:HAT-P-23b, Alam2022TheSystem, TheJWSTTransitingExoplanetCommunityEarlyReleaseScienceTeam2022IdentificationAtmosphere}. Hot Jupiters, especially those with inflated radii, are prime targets for transmission spectroscopy as they have large atmospheric scale heights due to their high temperatures, their hydrogen-dominated atmospheres and their low surface gravities. 
The sample of hot Jupiters studied to date exhibit a diverse range of atmospheric properties that can include: narrow or pressure-broadened sodium absorption \citep[e.g.][]{Fischer2016WASP-39b, Nikolov2018AnExoplanet, Alam2021EvidenceZone, McGruder2022ACCESS:Techniques}, detections of other atomic species and/or broad molecular bands \citep[e.g.][]{Lendl2017SignsWASP-103b,Carter2020DetectionWASP-6b, Ahrer2023EarlyNIRCam, Alderson2023EarlyG395H, Feinstein2023EarlyNIRISS, Rustamkulov2023EarlyPRISM}, Rayleigh scattering \citep[e.g.][]{Kirk2017RayleighHAT-P-18b, Chen2021AnWASP-104b} and sometimes super-Rayleigh slopes \citep[e.g.][]{Pont2013TheObservations,Alderson2020LRG-BEASTS:WASP-21b,Ahrer2022LRG-BEASTS:NTT/EFOSC2}, as well as high-altitude clouds muting absorption features \citep[e.g.][]{Gibson2013ANM, Knutson2014AGJ436b, Kreidberg2014Clouds1214b,Lendl2016FORS2WASP-49b,Louden2017AWASP-52b, Espinoza2019ACCESS:Magellan/IMACS, Spyratos2021TransmissionWASP-88b}. 

Transmission spectroscopy of hot Jupiters provides crucial information about the composition and chemistry of these exoplanets to understand their formation and migration process \citep[e.g.][]{Oberg2011TheAtmospheres, Madhusudhan2014TowardsMigration,Booth2017ChemicalDrift}, as well as what processes play a role in cloud and haze formation at these hot temperatures. The processes and parameters governing the presence or absence of clouds and hazes in the atmospheres of gas giants are still debated \citep[e.g.][]{Heng2016AWAVELENGTHS,Fu2017Planets, Fisher2018RetrievalDegeneracy, Pinhas2019H2OExoplanets, Gao2020AerosolHazes}. 

A larger sample size is needed to explore this parameter space, and the aim of the Low-Resolution Ground-Based Exoplanet Atmosphere Survey using Transmission Spectroscopy (LRG-BEASTS; ‘large beasts’) is to contribute to that by characterising a large number of gaseous exoplanets in transmission at optical wavelengths. This includes the detection of hazes, Rayleigh scattering and grey clouds in the atmospheres of WASP-52\,b \citep{Kirk2016TransmissionSurface,Louden2017AWASP-52b}, HAT-P-18\,b \citep{Kirk2017RayleighHAT-P-18b}, WASP-80\,b \citep{Kirk2018LRG-BEASTSWASP-80}, WASP-21\,b \citep{Alderson2020LRG-BEASTS:WASP-21b} and WASP-94A\,b \citep{Ahrer2022LRG-BEASTS:NTT/EFOSC2}, as well as detections of sodium absorption in the atmospheres of WASP-21\,b \citep{Alderson2020LRG-BEASTS:WASP-21b} and WASP-94A\,b \citep{Ahrer2022LRG-BEASTS:NTT/EFOSC2}. In addition, within LRG-BEASTS \citet{Kirk2019LRG-BEASTS:WASP-39b} analysed the atmosphere of WASP-39\,b revealing a supersolar metallicity and \citet{Kirk2021ACCESSWASP-103b} found tentative evidence for TiO in the atmosphere of the ultrahot Jupiter WASP-103\,b. 

In this paper we present the first transmission spectrum
of the exoplanet \mystar\,b.  Our observations were made using the EFOSC2 instrument on the New Technology Telescope (NTT) as part of the LRG-BEASTS survey. \mystar\,b was discovered within the HATSouth survey \citep{Bakos2013HATSouth:Telescopes1} by \citet{Brahm2018HATS-43bRange}. Their photometric observations, together with follow-up radial velocity measurements, confirm \mystar\,b, which orbits its G type host star in $4.74$\,days. Using TESS and Gaia data, \mystar\,b has been re-characterised by \citet{Louden2021HATS-34bGaia} who provided revised planetary and orbital parameters: \mystar\,b has a mass of $0.158\pm0.042$\,M$_{\textrm{Jup}}$ and a radius of $0.951\pm0.029$\,R$_{\textrm{Jup}}$, orbiting at a distance of $0.05272 \pm 0.00045$\,au; the equilibrium temperature was determined to $1082.1 \pm 8.2$\,K. Stellar and planet parameters are summarised in Table\,\ref{tab:stellar_planet_parameters}. The star \mystar\ does not appear to be very active as the RV measurements by \citet{Brahm2018HATS-43bRange} did not show any evidence for periodic modulation on a rotation period. 
Unfortunately, the signal-to-noise of the RV spectra was not sufficient to place constraints on the chromospheric activity from the Ca II H\&K lines \citep{Brahm2018HATS-43bRange}. The TESS light curves showed evidence for variability, with a possible period at around 15\,d, but if real this signal would also have been expected to be detected in the HATSouth light curve
\citep{Louden2021HATS-34bGaia}.

This paper is divided into the following sections. First, we describe the observations in Section\,\ref{sec:observations}, then discuss the data reduction and analysis in Sections\,\ref{sec:reduction} \& \ref{sec:analysis}. This is followed by our discussion and conclusions in Section\,\ref{sec:conclusions}.

\begin{table}
    \centering
    \caption{Parameters for the star \mystar\ and its planet \mystar\,b, with $V_{\textrm{mag}}$ and spectral type as determined by  \citet{Brahm2018HATS-43bRange} and all other parameters as revised by  \citet{Louden2021HATS-34bGaia}.} 
    \begin{tabular}{l c}
    \hline
        \multicolumn{1}{l}{\textbf{Parameter}} &  \multicolumn{1}{c}{\textbf{Value}} \\ \hline
         \multicolumn{2}{l}{Stellar parameters}  \\ \hline
        $V_{\textrm{mag}}$ & $13.634 \pm 0.050$ \\
        Spectral type & G \\
        Temperature $T_{\textrm{eff}}$ (K) & $5451 \pm 19 $ \\
        Age (Gyr) & $8.4\pm 1.9$   \\
        Surface gravity log $g$ (log$_{10}$(cm\,s$^{-2}$)) & $4.474 \pm 0.019$ \\
        Metallicity $[$Fe/H$]$ & $-0.029  \pm 0.039$  \\
        Mass ($M_\odot$) & $0.869 \pm 0.023$ \\
        Radius ($R_\odot$) & $0.894 \pm 0.010 $  \\
    \hline
    \multicolumn{2}{l}{Planetary parameters}  \\ \hline
    Period (d) & $4.7423749 \pm 0.0000043$ \\
    Mass ($M_{\textrm{Jup}}$) & $0.158\pm0.042$ \\
    Radius ($R_{\textrm{Jup}}$) & $0.951\pm0.029$ \\
    Semi-major axis (au) & $0.05272 \pm 0.00045$ \\
    Equilibrium temperature $T_{\textrm{eq}}$ (K) & $1082.1 \pm 8.2$ \\
    Inclination ($^\circ$) & $86.97\pm0.10$ \\
    Surface gravity log g (log$_{10}$(cm\,s$^{-2}$)) & $2.64 \pm 0.14$   \\ \hline
    \end{tabular}
    \label{tab:stellar_planet_parameters}
\end{table}

\section{Observations}
\label{sec:observations}

We observed \mystar\  with the NTT using the EFOSC2 instrument \citep{Buzzoni1984TheEFOSC} on the night of 17 August 2017\footnote{Based on observations collected at the European Southern Observatory under ESO programme 099.C-0390(A) (PI: Kirk).}. EFOSC2 is mounted at the Nasmyth B focus of the ESO NTT in La Silla, Chile, which has a Loral/Lesser CCD detector with a size of 2048 $\times$ 2048 pixels. The overall field of view is 4.1\,arcmin with a resolution of 0.12\,arcseconds per pixel and a pixel binning of $2\times2$ was applied. 

At our request, a slit with a width of 27\,arcsec was custom-built, with the aim of avoiding differential slit losses between target and comparison star. Grism \#13 was used for our spectroscopic measurements, providing a low-resolution ($R\sim 380$) spectrum from $3900 - 9000$\,\AA.

In total, 93 spectral frames were acquired, each with a relatively long exposure time of $240$\,s due to the relatively faint magnitude of both target and comparison star. The readout time was 22\,seconds. The observations were taken at an airmass ranging from 1.60 to 1.12 to 1.26. The illumination of the moon was at $16\%$ and it only rose towards the very end of the observation night at a distance to the target of $108^\circ$.

For calibration, 67 bias frames were acquired, as well as 112 flat frames (54 lamp, 53 sky, 5 dome) and 3 HeAr arc frames, taken at the beginning of the night. While we experimented with using flat frames in our data reduction, we did not use any in our final reduction as we found it to increase the noise in our data. This is in line with previous reports of similar analyses,  both by the LRG-BEASTS and ACCESS surveys \citep[e.g.][]{Rackham2017ACCESSPHOTOSPHERE, Bixel2019ACCESS:WASP-4b, Weaver2020ACCESS:K, Kirk2021ACCESSWASP-103b}.

A nearby star (UCAC4 169-000364) at a distance of 
1\,arcmin to the target star \mystar\ served as a comparison star and is not known to be a variable star. The two stars are a good match in both magnitude ($\Delta V_{\textrm{mag}}$ = 0.87) and colour ($\Delta (B-V) = 0.09$), thus well-suited for differential spectro-photometry.

\section{Data Reduction}
\label{sec:reduction}
LRG-BEASTS observations are commonly reduced using a custom-built \python\ pipeline, which is described in detail by  \citet{Kirk2018LRG-BEASTSWASP-80}. The data for \mystar\ have been reduced following this pipeline, but with modifications to the cosmic ray removal and wavelength calibration, introduced in \citet{Ahrer2022LRG-BEASTS:NTT/EFOSC2}. In the following we summarise the reduction steps. 

First, the biases were median-combined to produce a master bias. When executing the \python\ script for extracting the spectra from each science frame the master bias is subtracted from each science frame. %
However, before extracting the spectra from the individual frames, pixels affected by cosmic rays were identified and replaced with the median of the surrounding pixels. 

\begin{figure}
    \centering
    \includegraphics[width=\columnwidth]{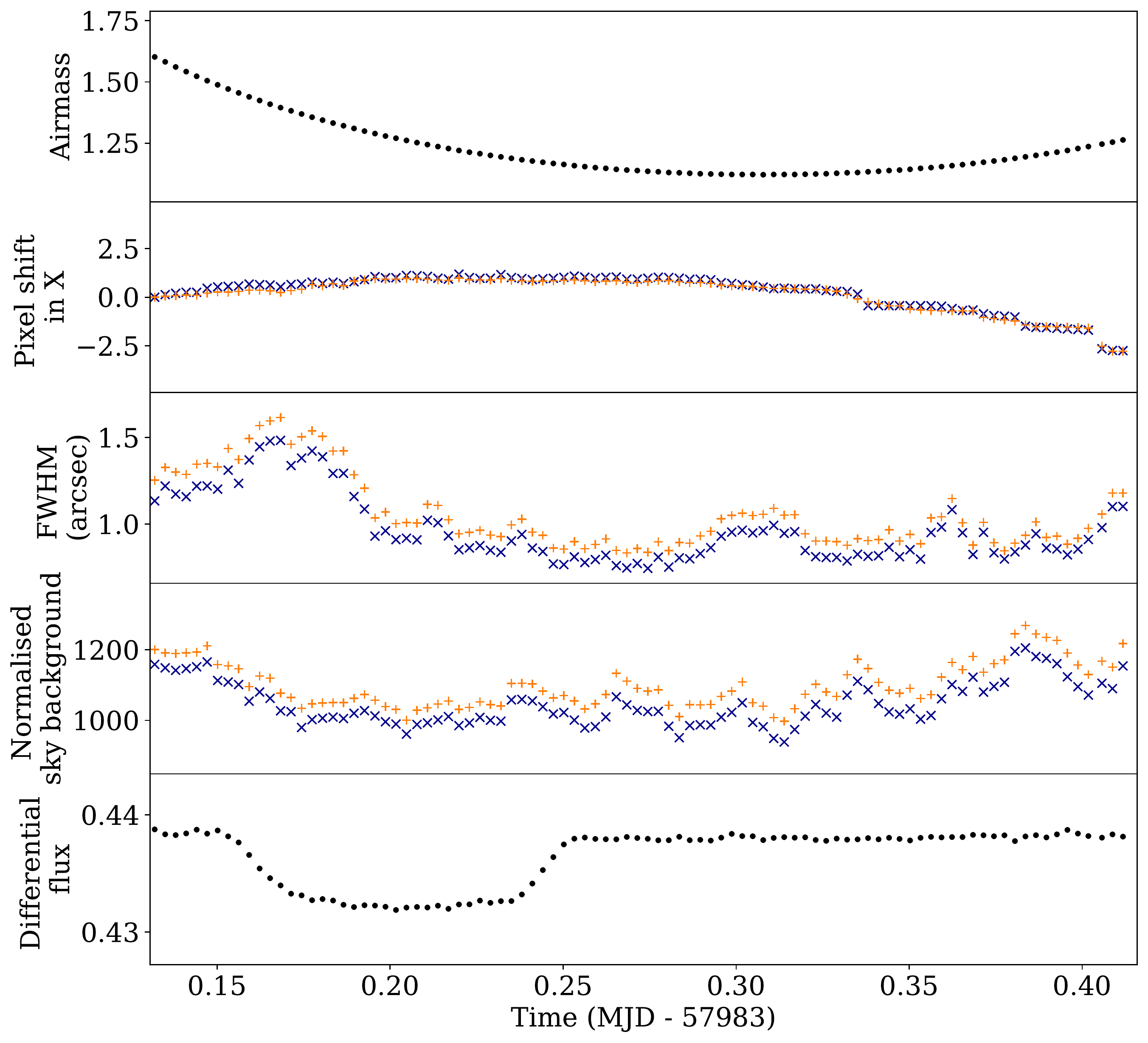}
    \caption{From top to bottom: variations of airmass, pixel shift along the X axis, FWHM, sky background and differential flux across the night. In the middle panels, the target is indicated with dark blue X symbols, and the comparison star with orange + symbols. }
    \label{fig:ancillary_plots}
\end{figure}

\begin{figure}
    \centering
    \includegraphics[width=\columnwidth]{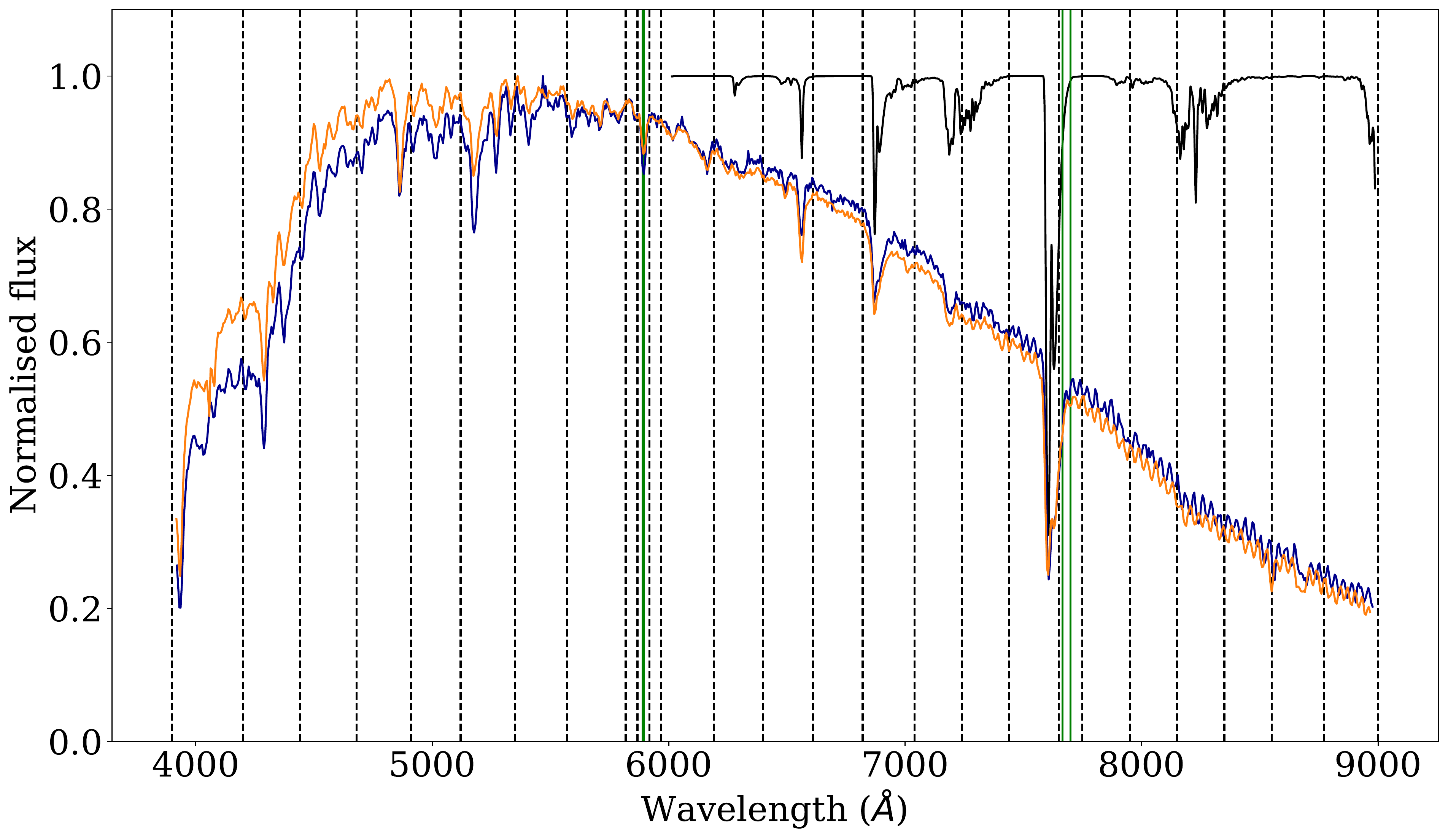}
    \caption{Normalised spectra of comparison star (orange) and target star (dark blue), as well as the expected strong telluric lines (black) in the redder part of the wavelength range. Wavelength bin edges are indicated with dashed black lines. Green lines indicate the position of the sodium doublet ($5890, 5895$\,\AA) and potassium doublet ($7665, 7699$\,\AA).}
    \label{fig:wvl_bins}
\end{figure}

An aperture width of 32\,pixels was applied to extract the spectral counts from each star. 
To fit the sky background we used a second order polynomial, which was fitted to regions of 50\,pixels either side of the stars at a distance of 5\,pixels from the edge of the aperture. Outliers of more than three standard deviations were masked from the fit. Extracted properties such as airmass, pixel shift along the slit, Full Width Half Maximum (FWHM), normalised sky background and differential white-light flux and their changes throughout the night are displayed in Fig.\,\ref{fig:ancillary_plots}. Example spectra are plotted in Fig.\,\ref{fig:wvl_bins}.  

Wavelength calibration follows the spectral extractions and is a two-step process. First, \textsc{RASCAL} \citep{Veitch-Michaelis2019RASCAL:Calibration} was utilised to find a wavelength solution using the HeAr arc frames. The second step is to optimise the wavelength calibration by fitting the positions of the stellar absorption lines in each frame, adjusting the solution, and then saving the wavelength solution for each frame individually. This allowed us to account for wavelength drifts between the frames throughout the night, which were of the order of $\sim 5$\,pixels or $\sim 20$\,\AA.

Lastly, the spectra were binned into 26 wavelength bins, computed by summing the flux within the corresponding wavelength range of each frame and dividing by the comparison star's flux in the same wavelength bin to correct for the affects of the Earth's atmosphere. Similarly, a white-light light curve was computed by defining one single bin across the whole wavelength range. Bin widths of $\sim 200$\,\AA\  (avoiding edges of strong stellar absorption lines) were applied across the whole spectral range, with the exception of two small ranges where we searched for  absorption by sodium and potassium, see Fig.\,\ref{fig:wvl_bins}.

Observations with EFOSC2 at wavelengths $> 7200$\,\AA\ are subject to fringing effects (see Fig.\,\ref{fig:wvl_bins}). We found that correcting for these effects in the individual spectra using flat fields was not possible as the fringing changed in amplitude and phase during the night and the acquired flat frames were taken before the observations started.

\section{Data Analysis}
\label{sec:analysis}

\subsection{Transit model}
Each transit light curve was described using the \textsc{batman} \python\ package \citep{Kreidberg2015BatmanPython} in combination with the analytic light curves from \citet{Mandel2002AnalyticSearches} and fitted using the nested sampling algorithm \textsc{PolyChord} \citep{Handley2015PolyChord:Sampling}. 
First, the white-light light curve was fitted using the ratio of planet to star radius $R_p/R_*$, the inclination of the system $i$, the scaled stellar radius $a/R_*$, the time of mid-transit $T_C$ and the two quadratic limb-darkening coefficients $u1$ and $u2$. We computed the limb-darkening coefficients with the Limb-Darkening Toolkit (\textsc{LDTk}) package \citep{Parviainen2015LDTK:Toolkit}, which uses \textsc{phoenix} models \citep{Husser2013ASpectra} based on the stellar parameters to determine $u1$ and $u2$ and their errors. One of them ($u2$) was held fixed to the generated value to avoid degeneracy,
while the other one was fitted for ($u1$) 
using a uniform prior with four times the generated error (see Table\,\ref{tab:WL_parameters}) to allow for small inconsistencies between the stellar model and the observation. This quadratic limb-darkening law provides a good fit to the data, see Section\,\ref{sec:light_curve_fitting}, and the fitted values for $u1$ were consistent with the model prediction. The Kipping parameterisation \citep{Kipping2013EfficientLaws} was also tested to check for potential effects in the transmission spectrum due to the chosen limb-darkening parameterisation, but we can confirm that this is not the case.

All priors for the system parameters can be found in Table\,\ref{tab:WL_parameters}, which were chosen to be uniform and wide ($\pm 5 \sigma$) centred on the previously reported literature values \citep[Table\,\ref{tab:stellar_planet_parameters};][]{Louden2021HATS-34bGaia}. Depending on the detrending method, additional parameters were added to the fitting (introduced in the following section).

\begin{table*}
    \centering
    \caption{Parameter values obtained from the white-light curve fitting and the respective priors. Values for semi-major axis $a$, radius of the star $R_*$ and radius of the planet $R_p$ and inclination $i$ are listed in Table~\ref{tab:stellar_planet_parameters}. The retrieved values for the parameters $a/R_*$, $i$ and $T_C$ listed here were fixed for the spectroscopic light curve fitting. }
    \label{tab:WL_parameters}
    \begin{tabular}{lcccc}
    \hline
    Parameter &  \multicolumn{2}{c}{Prior distribution and range} & Fitted values \\ \hline 
   Scaled stellar radius $a/R_*$ & Uniform & $a/R_* \pm  5\sigma_{a/R_*}$ &  $13.94^{+0.24}_{-0.65}$  \\
   Inclination $i$ ($^\circ$) & Uniform & $i$ $\pm$ 5$\sigma_{i}$ & $87.60^{+0.12}_{-0.33}$ \\
    Time of mid-transit $T_C$ (BJD) & Uniform & $0.9 \times\ T_C$, $1.1 \times\ T_C$ & $2457983.70725^{+0.00046}_{-0.00033}$  \\
    Transit depth $R_p/R_*$  & Uniform & $R_p/R_*$ $\pm$ 5$\sigma_{R_p/R_*}$& $0.11250^{+ 0.0.0018}_{-0.00083}$\\
    Limb-darkening coefficient $u1$ & Uniform & $u1 \pm 4\sigma_{u1}$ & $0.547\pm0.014$ \\
    Limb-darkening coefficient $u2$ & Fixed & -- & 0.1171 \\
    \hline
    \end{tabular} 
\end{table*}

The determined values for $a/R_*$, $i$ and $T_C$ from the white-light light curve fitting (Table\,\ref{tab:WL_parameters}) were then held fixed for the spectroscopic light curve fitting, which allowed us to fit for relative changes in transit depths over the wavelength range. Thus the fitting parameters for each of the 26 binned light curves were transit depth $R_p/R_*$, limb-darkening coefficient $u1$ and additional noise modelling parameters. 

\subsection{Light curve fitting}
\label{sec:light_curve_fitting}
\begin{figure}
    \centering
    \subfigure[Gaussian Process]{\includegraphics[width=\columnwidth]{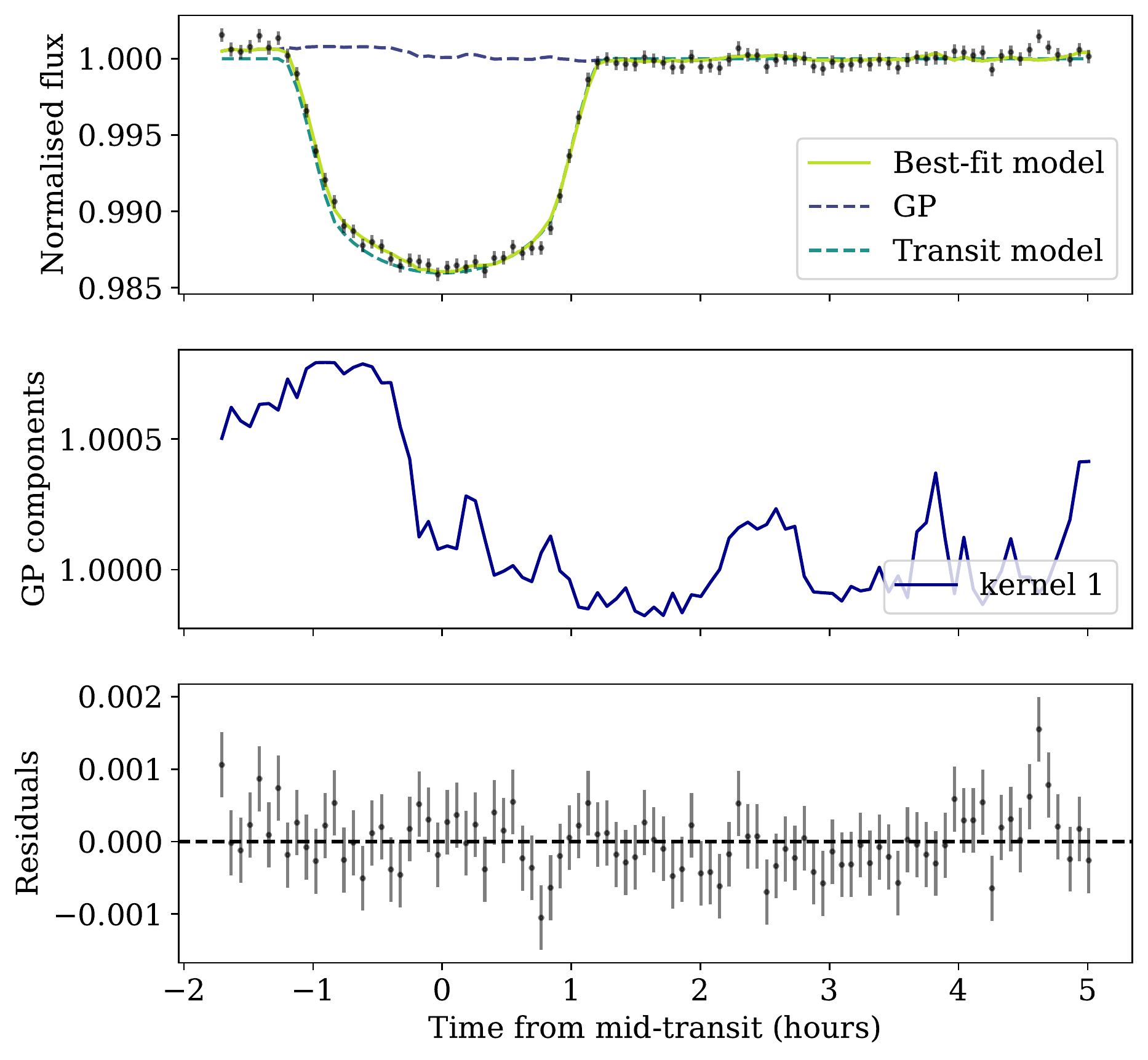}}
    \subfigure[Linear function of FWHM]{\includegraphics[width=\columnwidth]{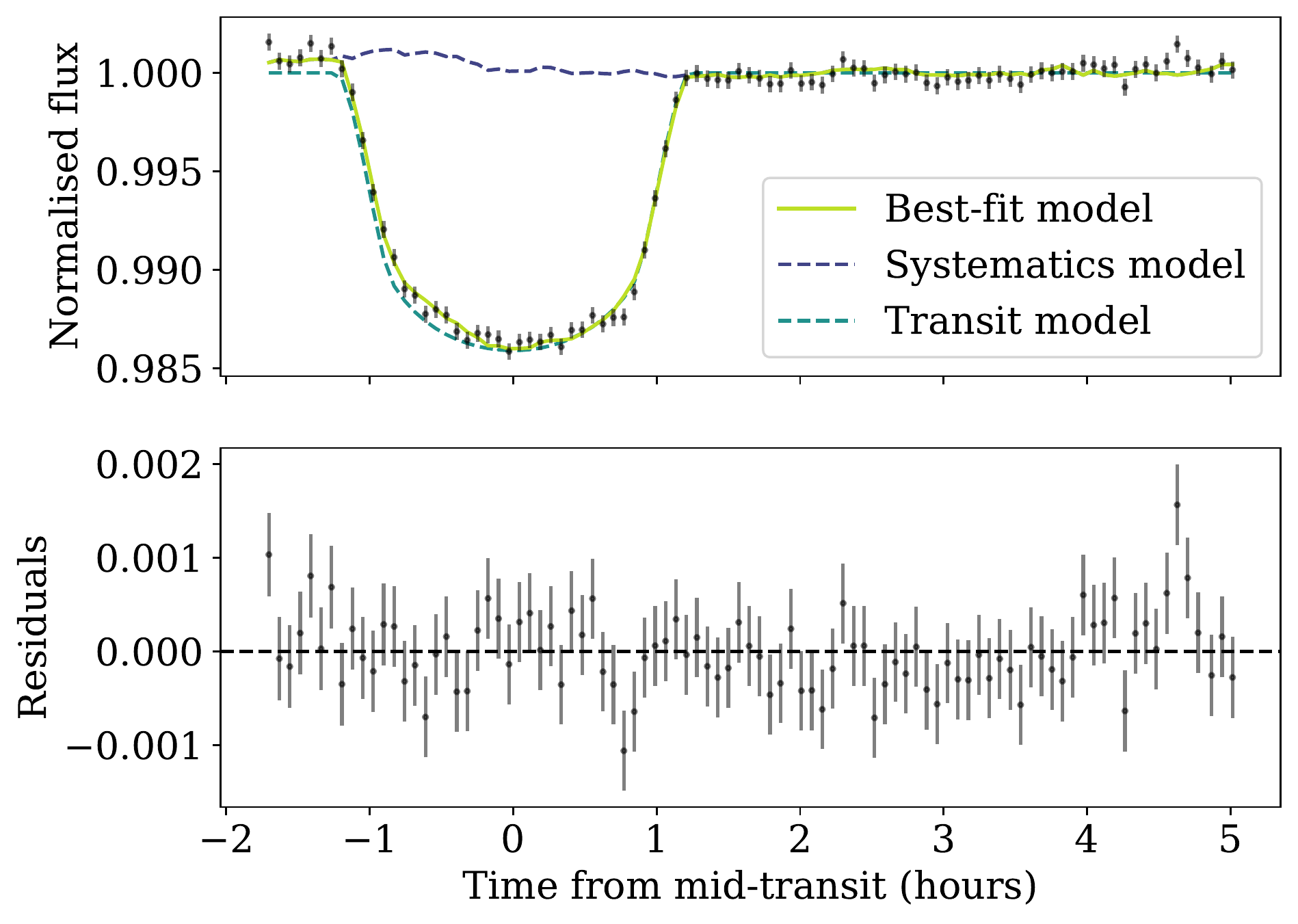}}
    \caption{
    The white-light light curve fitted with a transit and two different models to account for systematics. The best-fit model is plotted in green, while the individual components of the model are plotted in dashed turquoise for the transit model and dark blue for the respective systematics models. In the top panels, labelled (a), we use a GP model for systematics. 
    In the lower panels, labelled (b), we use a linear function of FWHM. 
    }    \label{fig:fitted_WL_curve}
\end{figure}

For detrending the white-light light curve, various different approaches were investigated e.g.\ different combinations of kernels and kernel inputs for a Gaussian Process (GP), 1st and 2nd order polynomials using airmass, FWHM, derotator angle, etc. However, all of these models retrieved very low amplitudes for their respective noise modelling, e.g.\ see amplitude of the best-fitting GP model in top panel in Fig.\,\ref{fig:fitted_WL_curve} which is $0.062$\,\% compared to the transit depth of $1.287\,$\%. In addition, the Bayesian evidence values for each of these fits did not statistically favour a particular GP model or parametric fitting model. The differences across all wavelengths in Bayesian evidences averaged at $0.5$ ($0.67\sigma$) and never exceeded $1$ ($< 1.15\sigma$). Consequently, we  opted to use only a 
linear dependence on the FWHM 
for detrending 
the white-light light curve, see bottom panel in Fig.\,\ref{fig:fitted_WL_curve}. 

To determine the transit depths for each wavelength bin, we fitted the individual light curves of the 26 bins with a transit model and a detrending model. We conducted an investigation of the systematics modelling, similar to the one done for the white-light light curve fit. This was to ensure that our transmission spectrum is independent of the choice of noise modelling, and to provide the best estimate of the uncertainties. 

\begin{figure*}
    \centering
    
    \includegraphics[width=\textwidth]{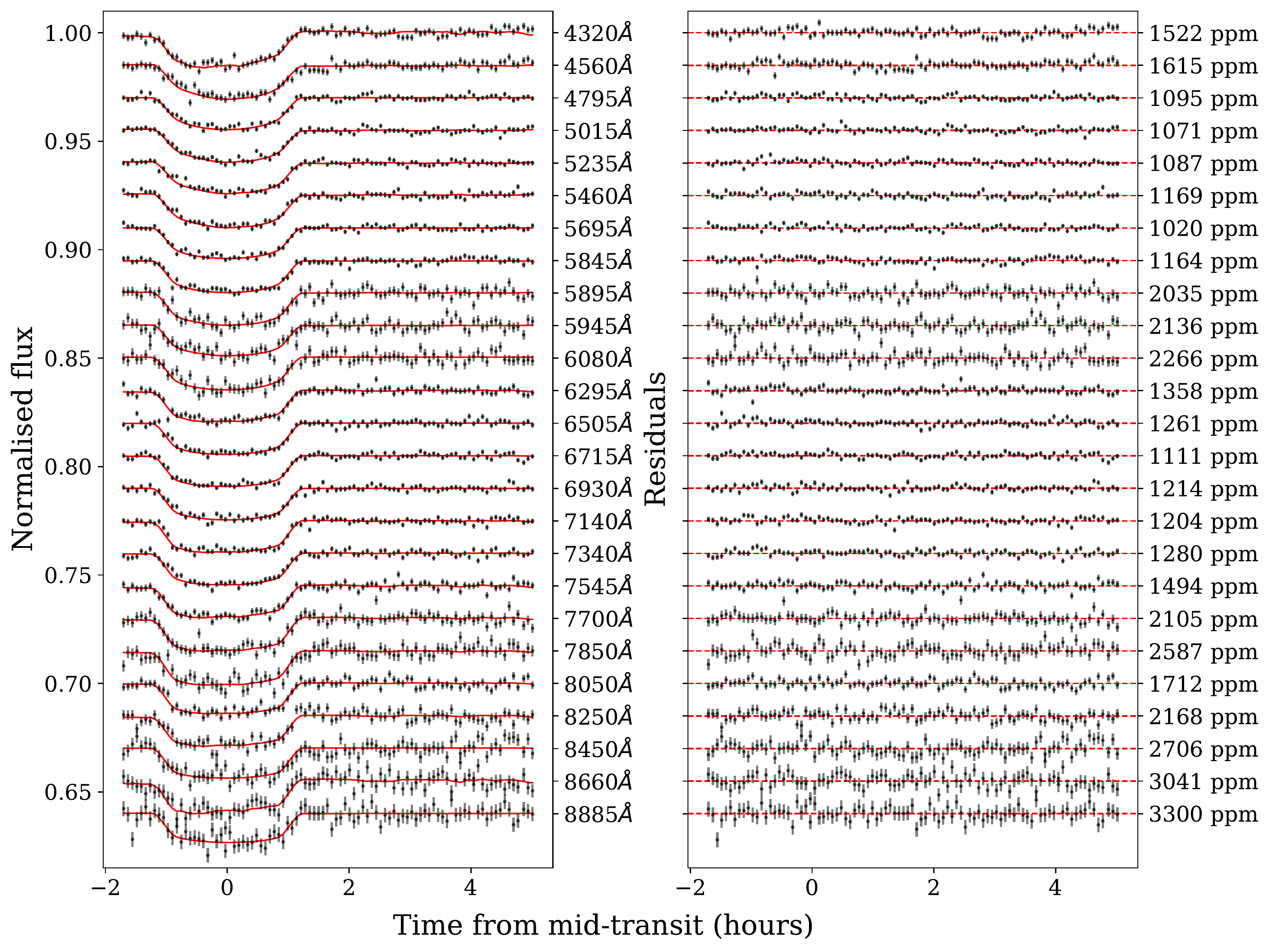}
    \caption{Left: Our fits (red) of the undetrended spectroscopic light curves (black) using a transit model and a linear in FWHM for detrending to the data with their respective centre wavelengths (blue end at the top) displayed on the right vertical axis. Right: Residuals of the corresponding light curve fitting. The scatter is quantified in the form of the RMS on the right vertical axis. }
    \label{fig:all_lightcurves_noGP}
\end{figure*}

The light curves show very little evidence for systematic trends such as drifts or correlated noise, see left panel in Fig.\,\ref{fig:all_lightcurves_noGP} for the raw light curves. We experimented with simple models to account for the small noise amplitudes, as well as using a transit model without any systematic modelling at all. First, linear models in time, airmass and FWHM were investigated, with the linear in FWHM performing the best according to the Bayesian evidence value of each spectroscopic light curve fit and an average fitted noise amplitude of $0.06\%$ or $600$\,ppm. In addition, we looked into GP models and sampled different types of kernels and kernel input, out of which the a exponential-squared model with FWHM as input resulted in the best choice, with an average fitted GP amplitude of $0.03\%$ or $300$\,ppm.  As both the linear in FWHM and GP model resulted in similar transit depths and small noise amplitudes, we chose the first, parametric model over the GP model due to its lower uncertainties in the transit depths. This results in an average precision of transit depth error equal to 1.03 $\times$ photon noise. The light curves and respective fits are shown in Fig.\,\ref{fig:all_lightcurves_noGP}, as well as the residual scatter of the fits and their respective Root Mean Square (RMS) values. 

The previously described models all favoured only small variations and FHWM as the detrending source for all spectroscopic bins. This led us to investigate using a common noise model \citep[e.g.\ as used in][]{Sing2012GTCSpectroscopy,Gibson2013TheFeatures,Lendl2016FORS2WASP-49b,Nikolov2016VLTGROUND,Nortmann2016TheHAT-P-32b,Huitson2017Gemini/GMOSWASP-4b,Todorov2019Ground-basedHAT-P-1b,Wilson2020Ground-basedWASP-103b,Kirk2021ACCESSWASP-103b, McGruder2022ACCESS:Techniques} in the hope of reducing our uncertainties and getting rid of common noise structures potentially dominating the systematics. In this method the GP component from the white-light light curve fit is subtracted from the spectroscopic light curves before fitting them individually.  However, this did not have the desired effect of improving the noise modelling and on average resulted in larger uncertainties. Therefore we did not pursue this method further. 

\begin{figure}
    \centering
    \includegraphics[width=\columnwidth]{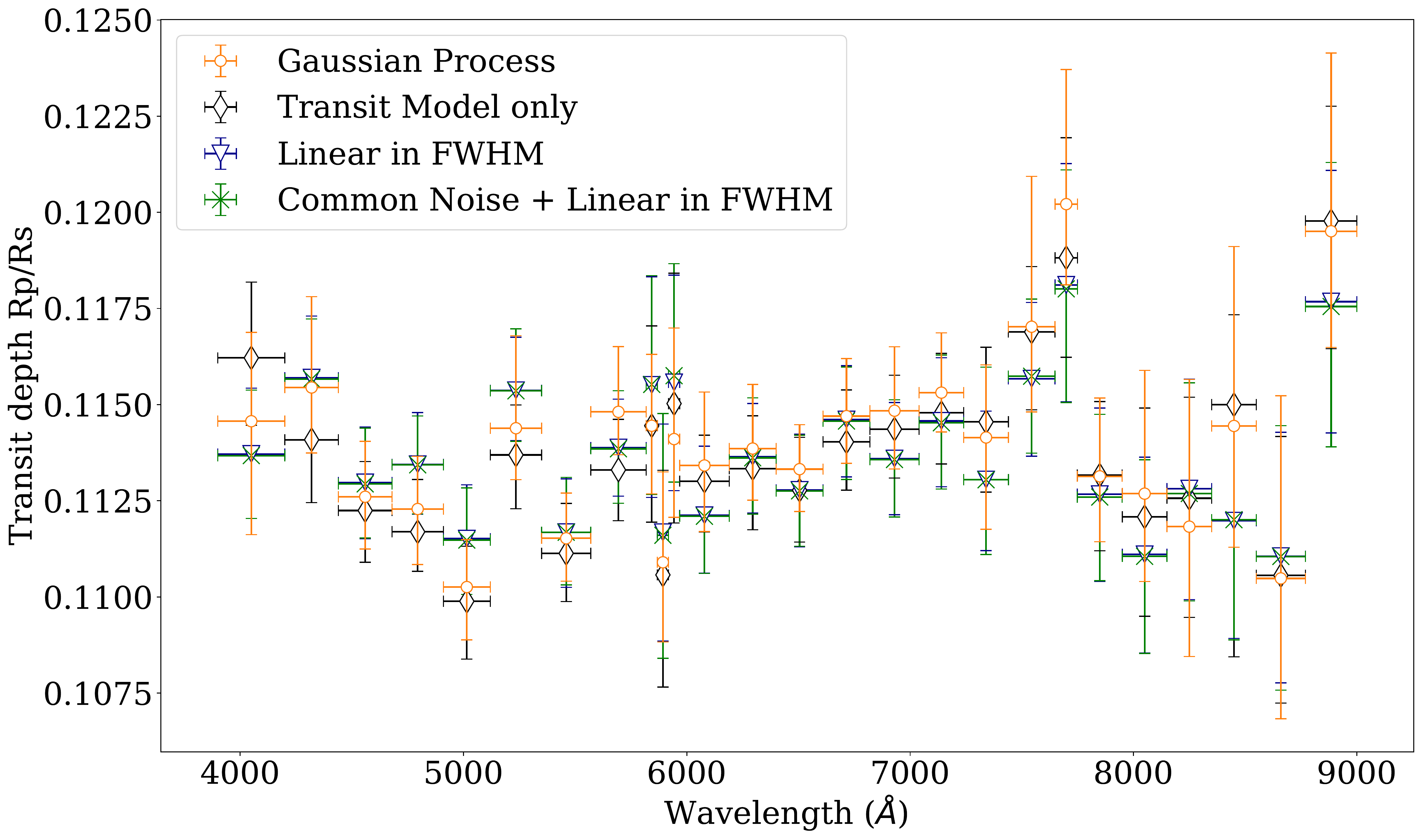}
    \caption{
    Transmission spectra of \mystar\,b using NTT/EFOSC2 observations. Median precisions of transit depths for $\sim 200$\,\AA\ wide bins are quoted in brackets in the description respectively. The orange and blue colours represent the resulting transmission spectrum using Gaussian Process (387\,ppm) and a linear in FWHM (357\,ppm) to account for systematics modelling, respectively. The black represents the case for when not using any noise modelling i.e.\ solely a transit model (326\,ppm). The green indicates a model where the GP component fitted to the white-light light curve was subtracted (common noise model) from the spectroscopic light curves and then a linear in FWHM was used to fit residual systematics (358\,ppm). The `Linear in FWHM' transmission spectrum is used for the retrieval analysis (see text for further details), but note that the bin centred on the potassium doublet (7665, 7699\,\AA) is not included as it is affected by the close-by strong telluric O$_2$ line. }
    \label{fig:transmission_spectrum_all}
\end{figure}

All computed transmission spectra using the GP model, the polynomial model, the common noise model and one without any detrending at all i.e.\ solely a transit model, are shown in Fig.\,\ref{fig:transmission_spectrum_all}. This demonstrates that our resulting transmission spectrum is independent of our choice of noise modelling. Following the points made above about each detrending approach, we selected a simple polynomial model, `Linear in FWHM', as the preferred detrending method. The final transmission spectrum in tabular form is displayed in Table\,\ref{tab:transmission_spectrum}.
Note that for our final spectrum we chose to dismiss the relatively large transit depth of the bin centred on the potassium doublet due the high chance of it being affected by the nearby strong telluric signal (O$_2$ A-band). Other studies in the past have come to similar conclusions when probing for potassium absorption with ground-based instruments \citep[e.g.][]{Kirk2017RayleighHAT-P-18b,McGruder2022ACCESS:Techniques}.

\begin{table}
    \centering
    \caption{Retrieved transmission spectrum of \mystar\,b in tabulated form using the 'Linear in FWHM' detrending approach, as plotted in Fig.~\ref{fig:transmission_spectrum_all}, excluding the bin centred on the K doublet.}
    \label{tab:transmission_spectrum}
    \begin{tabular}{cccc}
    \hline
        Bins (\AA)  & $R_p/R_*$ & u1 & u2 (fixed) \\ \hline
         3900 - 4200  & $0.1137^{+0.0018}_{-0.0017}$ & $0.92 \pm 0.02$ & -0.0737 \\
         4200 - 4440 & $0.1157\pm 0.0017$ & $0.87 \pm 0.02$ & -0.0523 \\
         4420 - 4680  & $0.1130\pm0.0015$ & $0.79 \pm 0.02$ & 0.0380\\
         4680 - 4910  &  $0.1134^{+0.0014}_{-0.0013}$ & $0.73 \pm 0.02$ &0.0726\\
         4910 - 5120  & $0.1115^{+0.0014}_{-0.0015}$ &  $0.71 \pm 0.02$ &0.0721\\
        5120 - 5350  & $0.1154\pm0.0014$ &  $0.67 \pm 0.02$& 0.0837\\
       5350 - 5570 & $0.1117^{+0.0014}_{-0.0015}$&  $0.63 \pm 0.01$ &0.1050\\
        5570 - 5818  & $0.1139\pm0.0013$ &  $0.59 \pm 0.01$&0.1241\\
        5818 - 5868  & $0.1155^{+ 0.0029}_{-0.0030}$  & $0.58 \pm 0.01$  &0.1330\\
        5868 - 5918 & $0.1117\pm0.0029$  & $0.58 \pm 0.01$  &0.1209\\
       5918 - 5968 & $0.1156^{+0.0028}_{-0.0029}$ &  $0.58 \pm 0.01$ &0.1295\\
        5968 - 6190 & $0.1121^{+0.0018}_{-0.0016}$  & $0.55 \pm 0.01$ & 0.1336\\
        6190 - 6400  & $0.1136^{+0.0014}_{-0.0015}$  & $0.53\pm 0.01$ &0.1364\\
        6400 - 6610  & $0.1128\pm0.0015$ & $0.50 \pm 0.01$ &0.1512\\
        6610 - 6820  & $0.1146\pm0.0015$  & $0.50\pm 0.01 $ & 0.1433\\
        6820 - 7040  & $0.1136\pm0.0015$  & $0.48\pm 0.01 $ &0.1446\\
        7040 - 7240 &  $0.1146^{+0.0017}_{-0.0018} $ & $0.47\pm 0.01$  &0.1449\\
        7240 - 7440&  $0.1130^{+0.0018}_{-0.0019} $ & $0.45\pm 0.01$ & 0.1452\\
       7440 - 7649 & $0.1157^{+0.0020}_{-0.0021}$ &  $0.44 \pm 0.01$ &0.1464\\
        7749 - 7950  & $0.1127\pm0.0023$  & $0.42 \pm 0.01$ & 0.147\\
        7950 - 8150  & $0.1111\pm0.0026$ & $0.42 \pm 0.01$ &0.1476\\
        8150 - 8350  & $0.1128\pm0.0029$  & $0.40\pm 0.01 $ & 0.1482 \\
        8350 - 8550  & $0.1120^{+0.0030}_{-0.0031}$  & $0.38\pm 0.01 $ &0.1474\\
        8550 - 8770 &  $0.1111\pm0.0033 $ & $0.37\pm 0.01$ & 0.1488\\
        8770 - 9000&  $0.1177\pm0.0035 $ & $0.37 \pm 0.01$ &0.1494\\
\hline
    \end{tabular}
\end{table}

\subsection{Atmospheric Retrieval}
\label{sec:retrieval}

We retrieve the transmission spectrum of \mystar\,b using the \textsc{HyDRA} \citep{Gandhi2018RetrievalHyDRA} and Aurora \citep{Welbanks2021Aurora:Spectra} atmospheric retrieval codes. Our model uses 14 free parameters which describe the atmospheric composition, thermal profile and cloud/haze properties (shown in Table~\ref{tab:priors}) to generate spectra of \mystar\,b to compare against the observations. We use high temperature molecular line lists to compute the cross sections and hence opacity for the spectrally active species, utilising the Kurucz line list for the atomic species Na and K \citep{Kurucz1995AtomicData}, and the ExoMol POKAZATEL line list for H$_2$O \citep{Tennyson2016TheAtmospheres,Polyansky2018ExoMolWater}. We spectrally broaden each line in the line list with both pressure and temperature, resulting in a Voigt profile \citep[see e.g.][]{Gandhi2020MolecularJupiters}. We also include collisionally induced absorption from H$_2$-H$_2$ and H$_2$-He interactions \citep{Richard2012NewCIA}, as well as Rayleigh scattering due to H$_2$. 

In addition to these sources of opacity we also include 4 free parameters to model and fit for a partially cloudy and/or hazy atmosphere, as any clouds/hazes can have a strong influence on the overall spectrum. We include a grey (wavelength independent) cloud deck, P$_\mathrm{cl}$, and two parameters which determine a wavelength dependent haze, with $\alpha_\mathrm{haze}$ the strength and $\gamma_\mathrm{haze}$ the wavelength dependence of the haze \citep[see e.g.,][]{Pinhas2018RetrievalAURA}. Finally, we include the cloud/haze fraction, $\phi_\mathrm{cl}$, as a free parameter, with the prior ranging from 0, representing a clear atmosphere, to 1, a fully cloudy/hazy atmosphere (see Table~\ref{tab:priors}).

We model the temperature profile of the atmosphere using the method described in \citet{Madhusudhan2009AAtmospheres}. This parametrisation breaks the atmosphere into three distinct layers, with the temperature at the top of the model atmosphere included as a free parameter. We also retrieve the transition pressures P$_1$ between the top layers 1 and 2 and P$_3$ between layers 2 and 3. The top two layers have temperature-pressure gradients $\alpha_1$ and $\alpha_2$ as free parameters. The final deepest layer of the atmosphere is fixed to an isotherm, and continuity of the temperature between these layers results in 6 free parameters for the temperature profile. We restrict our parametrisation to only allow non-inverted or isothermal temperature profiles given that we do not expect stratospheres for planets with such temperatures \citep[e.g.][]{Fortney2008AAtmospheres}, similar to previous work with transmission retrievals \citep[e.g.][]{Pinhas2019H2OExoplanets}. We also include an additional free parameter for the reference pressure, P$_\mathrm{ref}$, the point in the atmosphere where the radius of the planet is set. We model the atmosphere between 100-10$^{-6}$~bar with 100 layers evenly spaced in log pressure, and model the spectrum with 4000 wavelength points between 0.39-0.9~$\mu$m. Our Bayesian analysis is carried out using the Nested Sampling algorithm \textsc{MultiNest} \citep{Feroz2008MultimodalAnalyses, Feroz2009MULTINEST:Physics, Buchner2014X-rayCatalogue}. 

The retrieved constraints are shown in Table~\ref{tab:priors}, and the posterior distribution is shown in the Appendix, Fig.\,\ref{fig:posteriors}. For our retrievals we considered two competing scenarios: a cloudy/hazy atmosphere and a relatively-clear atmosphere. The first case, where clouds mask atomic and molecular species in the transmission spectrum of \mystar\,b, is statistically preferred to 3.0$\sigma$ due to the relatively featureless spectrum, when using Bayesian model evidence comparisons \citep[e.g.,][]{Benneke2013HowSuper-earths, Welbanks2021Aurora:Spectra}.
In the alternative, less statistically preferred scenario of a clear atmosphere, where clouds do not mask the atomic and molecular species, we can place constraints on the abundance of K and Na. There is no visible feature of Na in the spectrum, hence we place an upper limit on Na abundance of $\log(\rm{Na}) < -4.45$ to 3$\sigma$, i.e., less than $20 \times$solar Na abundance for this cloud-free scenario.
This is a conservative upper limit, since the lack of features in the transmission spectrum drives the atmospheric temperatures in the model to the lower end of the prior, which decreases the atmospheric scale height and thereby the strength of features. There is therefore a degeneracy between temperature and abundance, and an atmospheric temperature closer to the equilibrium temperature would give a tighter limit on abundance.  

Additionally, we assess the impact of unnoculted star spots and faculae in the transmission spectrum of \mystar\,b using \textsc{Aurora} \citep{Welbanks2021Aurora:Spectra}. We allow for the possibility of a contaminated stellar photosphere and retrieve for three additional parameters to the fiducial model described above. These are, the photospheric temperature (Gaussian prior centred at effective temperature of the star and a width of 100 K), the fraction of unnoculted spots or faculae (uniform prior between 0 and 50 \%), and the temperature of these inhomogeneities (uniform prior from 0.5 to 1.5 times the effective temperature of the star). Priors are in line with what is recommended by \citet{Pinhas2018RetrievalAURA}.  The retrieved properties stellar properties are in agreement with the possibility of a spotless star. The retrieved photospheric temperature of \mystar\ is consistent with the reported value in Table \ref{tab:stellar_planet_parameters}, with a relatively low fraction of spots (i.e., $2\sigma$ upper limit of $\lesssim 22\%$) with temperatures consistent with the photospheric stellar temperature at $2\sigma$. The presence of stellar heterogeneities is not preferred since its Bayesian evidence value is lower relative to our fiducial model. Based on these observations and the models considered here, we find no evidence for stellar contamination affecting our observations. 


\begin{table}
    \centering
    \def\arraystretch{1.5}
    \begin{tabular}{c|c|c}
    \hline
\textbf{Parameter}              & \textbf{Prior Range} & \textbf{Retrieval Constraint}\\
\hline
$\log(X_\mathrm{H_2O})$  & -15 $\rightarrow$ -1 & $-8.4^{+4.8}_{-4.2}$\\
$\log(X_\mathrm{Na})$ & -15 $\rightarrow$ -1 & $-10.1^{+3.5}_{-3.0}$\\
$\log(X_\mathrm{K})$ & -15 $\rightarrow$ -1 & $-8.6^{+3.3}_{-4.0}$\\
$T_\mathrm{top}$ / K & 750 $\rightarrow$ 2500 & $1167^{+530}_{-300}$\\
$\alpha_1\, /\, \mathrm{K}^{-\frac{1}{2}}$ & 0 $\rightarrow$ 1 & $0.67^{+0.21}_{-0.23}$\\
$\alpha_2\, /\, \mathrm{K}^{-\frac{1}{2}}$ & 0 $\rightarrow$ 1 & $0.61^{+0.25}_{-0.27}$\\
$\log(P_1 / \mathrm{bar})$ & -6 $\rightarrow$ 2 & $-1.7\pm 1.7$\\
$\log(P_2 / \mathrm{bar})$ & -6 $\rightarrow$ 2 & $-4.1^{+1.6}_{-1.3}$\\
$\log(P_3 / \mathrm{bar})$ & -2 $\rightarrow$ 2 & $0.60^{+0.90}_{-1.35}$\\
$\log(P_\mathrm{ref} / \mathrm{bar})$ & -4 $\rightarrow$ 2 & $-2.51^{+1.02}_{-0.86}$\\
$\log(\alpha_\mathrm{haze})$ & -4 $\rightarrow$ 6 & $-0.0^{+2.8}_{-2.5}$\\
$\gamma_\mathrm{haze}$ & -20 $\rightarrow$ -1 & $-11.3^{+6.3}_{-5.5}$\\
$\log(P_\mathrm{cl}/\mathrm{bar})$ & -6 $\rightarrow$ 2 & $-4.42^{+1.24}_{-0.94}$\\
$\phi_\mathrm{cl}$ & 0 $\rightarrow$ 1 & $0.79^{+0.13}_{-0.19}$\\ \hline
    \end{tabular}
    \caption{Parameters and uniform prior ranges for our retrieval. We retrieve the Na, K and H$_2$O abundances, temperature profile, and partial cloud/haze parameters. Our temperature profile includes 6 free parameters, and our cloud/haze parametrisation includes 4 free parameters (see Section\,\ref{sec:retrieval}).}
    \label{tab:priors}
\end{table}

\begin{figure*}
    \centering
    \includegraphics[width=\textwidth]{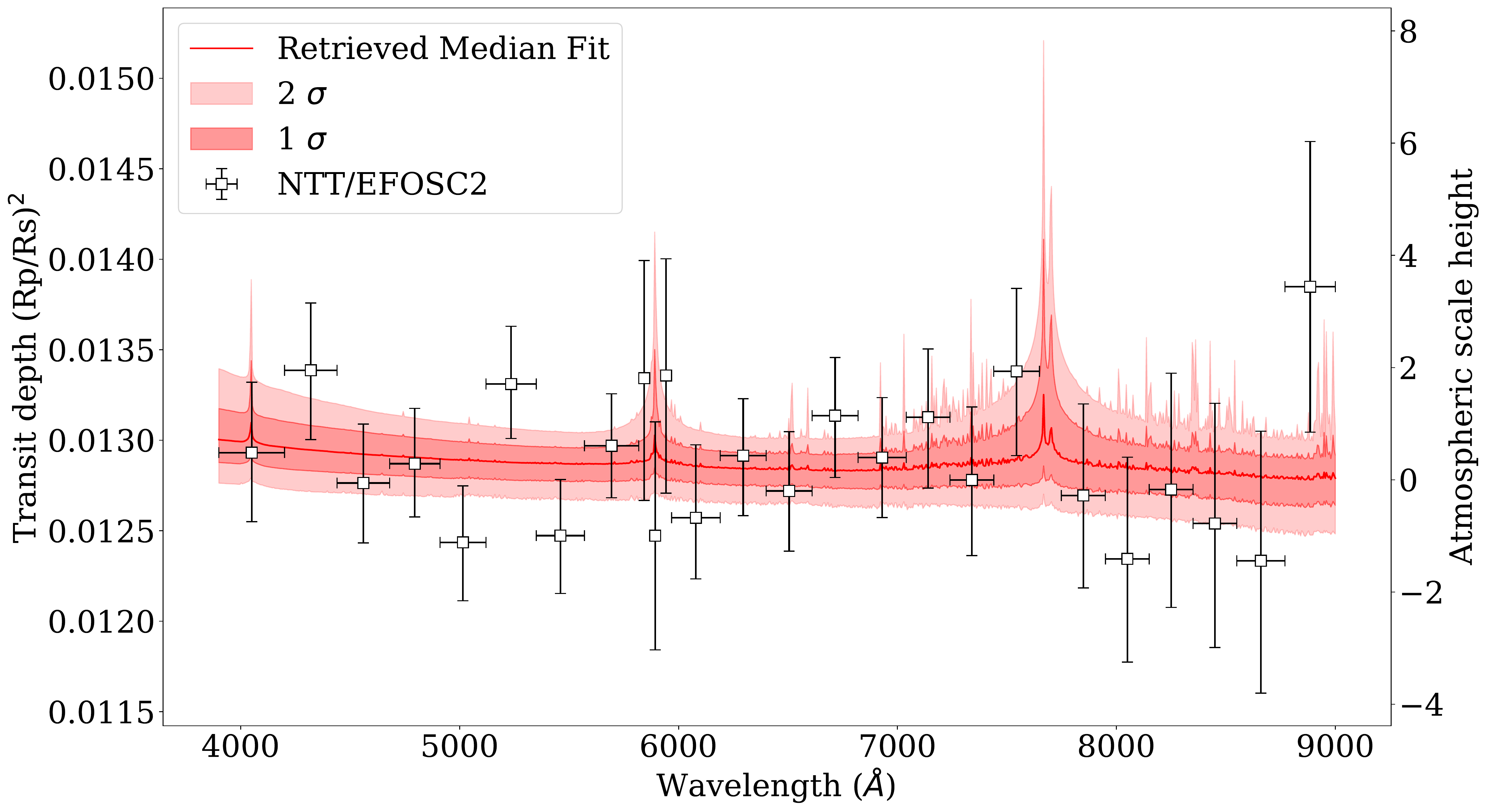}
    \caption{Transmission spectrum of \mystar\,b as observed by NTT/EFOSC2 and using linear in FWHM detrending (black), and the median retrieved atmospheric model (red), including the respective $1\sigma$ and $2\sigma$ confidence intervals. It is shown that the retrieved transmission spectrum is relatively featureless, suggesting high-altitude clouds in the atmosphere. Note that narrower bins around the Na doublet ($5890, 5895$\,\AA) are used to probe for absorption and the bin centred on the K doublet ($7665, 7699$\,\AA) was disregarded due to the close strong O$_2$ telluric line.  }
    \label{fig:transmission_spectrum_with_model}
\end{figure*}

\section{Discussion \& Conclusions}
\label{sec:conclusions}
We presented the analysis and results of spectroscopic NTT/EFOSC2 data of \mystar\ b in transmission. The inflated, Jupiter-sized exoplanet orbits its relatively faint (V$_\textrm{mag}$ = 13.6) G type host star in a 4.7-day period and has an equilibrium temperature of 1100\,K \citep{Louden2021HATS-34bGaia}. 

One transit was observed with NTT/EFOSC2 using the method of long-slit spectroscopy and a comparison star was used to conduct differential spectroscopy. A total of 93 spectral frames with exposure times of $240$\,s were acquired. The resulting light curves did not show  noise structures beyond a weak dependence on seeing, with fitted average amplitudes of $600$\,ppm for our best noise model, which included a linear detrend against FWHM.

We extracted the transmission spectrum in 26\,bins, covering the wavelength range of $3900 - 9000$\,\AA\ with a median transit depth uncertainty of $357$\,ppm for the $\sim200$\AA\ wide bins. 
The measured transmission spectrum is relatively featureless, it does not show a sodium feature or a scattering slope. The fitted, relatively large transit depth at the wavelength of the potassium doublet was dismissed as an effect of the nearby strong telluric signal due to the O$_2$ A-band. 
Our atmospheric retrieval analysis of the transmission spectrum of \mystar\,b favours a cloudy atmosphere with $3.0\sigma$ confidence. In an alternative cloud-free model we place a conservative  upper limit on the Na abundance of $20\times$solar ($3\sigma$ confidence). 
Including stellar activity in our retrievals results in lower Bayesian evidence and no meaningful constraints on the additional parameters. 
If activity were to play a role in the shape in our transmission spectrum, we would expect to retrieve constraints on the spot coverage fraction or temperature of the spots. 
Thus the cloudy atmosphere model without the additional stellar activity parameters is favoured.


\section*{Acknowledgements}
This research has made use of the NASA Exoplanet Archive, which is operated by the California Institute of Technology, under contract with the National Aeronautics and Space Administration under the Exoplanet Exploration Program. PJW acknowledges support from STFC under consolidated grants ST/P000495/1 and ST/T000406/1. SG is grateful to Leiden Observatory at Leiden University for the award of the Oort Fellowship. JK acknowledges financial support from Imperial College London through an Imperial College Research Fellowship grant.

\section*{Data Availability}
The raw data used in our analysis are available from the ESO data archive under ESO programme 099.C-0390(A) (PI: Kirk). The reduced light curves presented in this article will be available via VizieR at CDS \citep{Ochsenbein2000TheCatalogues}. 



\bibliographystyle{mnras}




\appendix

\section{Posterior distributions}
\begin{figure*}
\centering
	\includegraphics[width=\textwidth,trim={0cm 0cm 0cm 0},clip]{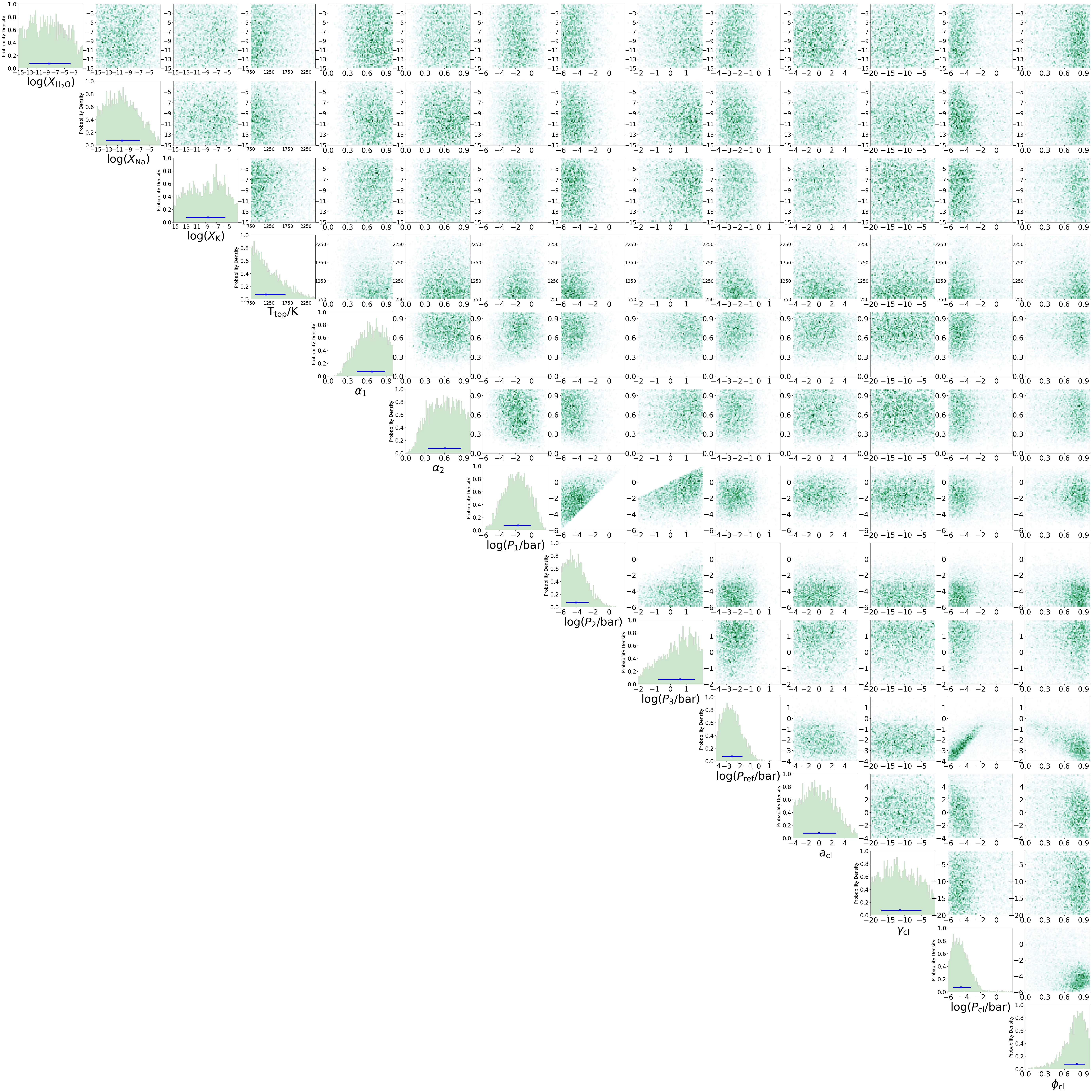}
    \caption{Posterior distribution of \mystar\,b from the retrieval of the NTT/EFOSC2 observations of \mystar\,b. We retrieved three chemical species, H$_2$O, Na and K, and parametrised the atmospheric temperature profile with six parameters, as discussed in Section\,\ref{sec:retrieval}. We also include additional parameters for the reference pressure and partial clouds/hazes.}     
\label{fig:posteriors}
\end{figure*}


\bsp	
\label{lastpage}
\end{document}